\documentclass[journal,twoside,web]{ieeecolor}

\usepackage{amsmath,amssymb,amsfonts}
\usepackage{algorithmic}
\usepackage{graphicx}
\usepackage{textcomp}
\usepackage{hyperref}
\usepackage{caption}
\usepackage{generic}

\def\BibTeX{{\rm B\kern-.05em{\sc i\kern-.025em b}\kern-.08em
    T\kern-.1667em\lower.7ex\hbox{E}\kern-.125emX}}

\begin{document}

\title{Predicting speech intelligibility from EEG in a non-linear classification paradigm}

\author{Bernd Accou, Mohammad Jalilpour Monesi, Hugo Van hamme and Tom Francart
\thanks{
Submitted for review on 7 June 2021. The research conducted in this paper is funded by KU Leuven Special Research Fund C24/18/099 (C2 project to Tom Francart and Hugo Van hamme), by a Ph.D. grant (1S89620N) of the Research Foundation Flanders (FWO) and from the European Research Council (ERC) under the European Union’s Horizon 2020 research and innovation program (grant agreement No 637424, ERC starting Grant to Tom Francart)
}
\thanks{B. Accou and M. J. Monesi are with the Department of Neuroscience and the Department of Electrical Engineering of KU Leuven, Belgium (email: bernd.accou@kuleuven.be, mohammad.jalilpourmonesi@esat.kuleuven.be)}
\thanks{H. Van hamme is with the Department of Electrical Engineering of KU Leuven, Belgium (email: hugo.vanhamme@esat.kuleuven.be)}
\thanks{T. Francart is with the Department of Neuroscience of KU Leuven, Belgium (email: tom.francart@kuleuven.be)}
}

\maketitle

\begin{abstract}
Objective: Currently, only behavioral speech understanding tests are available, which require active participation of the person being tested. As this is infeasible for certain populations, an objective measure of speech intelligibility is required. Recently, brain imaging data has been used to establish a relationship between stimulus and brain response. Linear models have been successfully linked to speech intelligibility but require per-subject training. We present a deep-learning-based model incorporating dilated convolutions that operates in a match/mismatch paradigm. The accuracy of the model's match/mismatch predictions can be used as a proxy for speech intelligibility without subject-specific (re)training.
Approach: We evaluated the performance of the model as a function of input segment length, EEG frequency band and receptive field size while comparing it to multiple baseline models. Next, we evaluated performance on held-out data and finetuning. Finally, we established a link between the accuracy of our model and the state-of-the-art behavioral MATRIX test.
Main results: The dilated convolutional model significantly outperformed the baseline models for every input segment length, for all EEG frequency bands except the delta and theta band, and receptive field sizes between 250 and 500~ms. Additionally, finetuning significantly increased the accuracy on a held-out dataset. Finally, a significant correlation (r=0.59, p=0.0154) was found between the speech reception threshold estimated using the behavioral MATRIX test and our objective method.
Significance: Our method is the first to predict the speech reception threshold from EEG for unseen subjects, contributing to objective measures of speech intelligibility.

\end{abstract}


\section{Introduction}
\label{sec:intro}

Current tests to diagnose hearing loss require the active participation of the person being tested. This can be labor and time-intensive in certain populations or even impossible in others (e.g., young children). Furthermore, most tests use artificial stimuli such as tones or clicks, which are not representative of real-world hearing. Therefore, there is a need for an objective and automatic measure of speech intelligibility with more ecologically valid stimuli.

Recently, an objective measure of speech intelligibility has been proposed using EEG or MEG data, based on a measure of cortical tracking of the speech envelope \cite{ding_emergence_2012, vanthornhout_speech_2018, lesenfants_data-driven_2019}. Tracking can be measured with 3 groups of models: backward models \cite{crosse_multivariate_2016, vanthornhout_speech_2018, Gillis2021.01.21.427550}, forward models \cite{crosse_multivariate_2016, lesenfants_data-driven_2019} and hybrid models \cite{de_cheveigne_decoding_2018, wong_accurate_2019}.  While the results for backward and forward models are promising and can be linked to speech understanding \cite{vanthornhout_speech_2018, akbari_towards_2019, iotzov_eeg_2019, lesenfants_predicting_2019}, the variability for repeated measurements is high and the correlation between the reconstructed and stimulus envelope remains low \cite{verschueren_neural_2019}. Subject-specific models are more commonly used than subject-independent models. However, subject-independent models are more attractive from an application perspective as no training data needs to be recorded for evaluation purposes.

A possible improvement is moving to a non-linear model\cite{ciccarelli_comparison_2018}, which is better equipped to model the brain, a highly complex and non-linear system, across subjects. For instance, it has been shown by Ding et al. \cite{ding_emergence_2012} that depending on the level of attention and state of arousal of the subject, response latencies can change dramatically, which cannot be modeled using a purely linear approach. For intracranial electrodes \cite{akbari_towards_2019, yang_speech_2015}, better results have been achieved with simple artificial neural networks compared to linear models for the same modality.
 
For EEG,  convolutional networks have been applied for auditory attention detection \cite{deckers_eeg-based_2018, ciccarelli_comparison_2018, geirnaert_neuro-steered_2020, taillez_machine_2017}. Instead of a two-step approach (reconstructing the attended stimulus and comparing the similarity with the actual stimuli), these convolution-based models can classify the attended speaker directly from the EEG and envelope of the speech signal. This approach is highly successful, as auditory attention can be decoded in 10 seconds with 81\% median accuracy \cite{ciccarelli_comparison_2018} and the locus of attention in 1-2 seconds with 80\% accuracy \cite{deckers_eeg-based_2018}, and has been shown to outperform models that perform classification based on an intermediate measure of similarity/correlation \cite{ciccarelli_comparison_2018}. Additionally, O'Sullivan et al. \cite{osullivan_attentional_2015} showed that decoding results correlated with a behavioral measure (responses to multiple-choice questions).

Inspired by the recent developments in auditory attention decoding and CCA\cite{de_cheveigne_decoding_2018}, we introduced a match/mismatch paradigm \cite{10.1088/1741-2552/abf771} in \cite{accou_modeling_2020}, based on the method of de Cheveigné et al. \cite{10.1088/1741-2552/abf771}, to relate an acoustic stimulus to a corresponding EEG recording. In this paradigm, a model with 3 inputs is presented: (a segment of) EEG, the speech stimulus envelope and an imposter envelope. The task of the model in this paradigm is to determine which of the input envelope segments correspond to the EEG segment. We showed in \cite{accou_modeling_2020} that this approach yields relatively high performance for short envelope segments ($\sim 90\%$).

We propose a dilated convolutional network as the basis of an objective measure of speech intelligibility (in our case, word recognition accuracy in noise). Dilated convolutions are a constrained way to do convolutions, with fewer weights, as used in WaveNet \cite{oord_wavenet:_2016}. By eliminating redundancy and leaving holes in subsequent convolutional layers, each output node can obtain information from exponentially more input nodes. At the same time, the number of weights increases linearly per layer instead of exponentially. 
This network is trained in a subject-independent manner in the match/mismatch paradigm. In section \ref{sec:mm_experiments} of this paper, we evaluate our proposed model \cite{accou_modeling_2020} compare it to a baseline. In section \ref{sec:beh_experiments}, we show that our model can be used to estimate the speech intelligibility of unseen subjects by fitting a sigmoid function on the accuracy of the model predictions, based on the technique used by Vanthornhout et al \cite{vanthornhout_speech_2018}. In contrast with Vanthornhout et al \cite{vanthornhout_speech_2018}, however, we use a subject-independent model (dilated convolutional model) instead of a subject-specific linear decoder.

\begin{figure}
	\centering
	\includegraphics[width=0.49\textwidth]{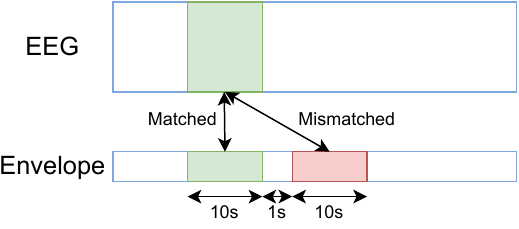}
	\caption{To ensure similarity to the matched speech envelope segment, the imposter speech envelope segment is extracted from the same speech recording, 1 second in the future from the time-aligned speech envelope segment.}
	\label{fig:matchmismatch}
\end{figure}

\section{Methods}
\label{sec:methods}

\subsection{Datasets}
In this paper, 2 datasets are used: our own collected dataset (Fairytales/held-out) and a subset of the dataset used by Vanthornhout et al. \cite{vanthornhout_speech_2018} (MatrixEEG).

\subsubsection{Fairytales and held-out dataset} 
\label{sec:fairytales}
For this dataset, 68 subjects between 18-30 years old were recruited. This study was approved by the Medical Ethics Committee UZ KU Leuven/Research (KU Leuven, Belgium) with reference S57102 and all participants provided informed consent. To ensure that participants had normal hearing, they were subjected to pure-tone audiometry and an adaptive MATRIX test in Flemish \cite{luts_development_2014}.  For the pure-tone audiometry, normal hearing was defined as having all hearing thresholds $\leq$ 25 dBHL. The MATRIX test consisted of 3 trials (2 for training purposes, 1 for the actual testing) in which 20 sentences (spoken by a female voice in Flemish) were presented to the subject binaurally at 62~dBA. Each sentence consisted of 5 words (proper name-verb-number-color-noun) and carried little to no semantic meaning. During a trial, the signal-to-noise ratio was adjusted according to the score the subject obtained on the previous sentence, converging to the point where the subject understands approximately 50\% of all words, known as the speech reception threshold (SRT). Subjects with SRT's higher than -3dB were excluded.

Subsequently, all subjects listened to fairy tales narrated in Flemish (without noise) while their EEG was recorded. The recordings were randomly selected for each subject from a pool of 10 stories. All recordings were approximately the same length (14 minutes and 29 seconds ± 1 minute and 7 seconds), and their presentation order was randomized across subjects. To motivate subjects to pay attention during listening, they were notified beforehand that a question would be asked about the story's content after each recording. Additionally, subjects were given 3 breaks throughout the recording session.
This dataset is split into 2 parts for our experiments: the Fairytales dataset and the held-out dataset. Both of these datasets will be used for the training of the dilated convolutional model (section \ref{sec:mm_experiments}).
The Fairytales dataset contains 48 subjects. Of these 48 subjects, 23, 20, 4 and 1 subjects listened to 8, 7, 6 and 2 stimuli, respectively, accumulating to approximately 80 hours of data (64 hours for the train set, 8 hours for the validation and test set). The 20 remaining subjects in the held-out dataset all listened to 8 recordings, accumulating approximately 36.5 hours of data (29.5 hours for the train set, 3.5 hours for the validation and test set).

\subsubsection{MatrixEEG dataset}
\label{sec:matrix_eeg}

For the speech intelligibility estimation part of this paper, a subset of the dataset described by Vanthornhout et al. \cite{vanthornhout_speech_2018} and Lesenfants et al. \cite{ lesenfants_predicting_2019} is used. This dataset consists of 20 young normal hearing subjects who were tested behaviorally and objectively using EEG. For the behavioral part, the SRT was determined using a constant MATRIX test in Flemish, which is considered the gold standard in behavioral testing \cite{decruy_self-assessed_2018}. In contrast with the adaptive MATRIX test, used in the screening procedure of the Fairytales dataset \ref{sec:fairytales}, the SNR is constant throughout an entire list of sentences and does not depend on the previous answer of the subject. Further details about the behavioral testing are specified by Lesenfants et al. \cite{lesenfants_predicting_2019}.

Next, all subjects listened to MATRIX lists at 7 SNR's (-12.5, -9.5, -6.5, -3.5, -0.5, 2.5, no noise) while their EEG was recorded. For each SNR, 40 sentences were presented in random order, while silences between sentences ranged from 0.8 to 1.2 seconds. This was repeated 2 times to evaluate test-retest reliability. After each SNR presentation, subjects were asked a question about the sentences (e.g., "What color were the boats?") to motivate them to pay sufficient attention. This dataset was only used to evaluate if the SRT (as found by the MATRIX test) can be estimated based on the performance of the dilated convolutional model. Each of the subjects in the MatrixEEG dataset also listened to the fairytale Milan (which is also present in the Fairytales dataset). This data was only used for fine-tuning of the dilated convolutional model to improve SRT estimation in Section \ref{sec:beh_vs_obj}.

\subsection{Preprocessing}
\label{sec:prep}
A BioSemi Active Two system with 64 active electrodes and 2 extra mastoid electrodes was used to record EEG at a sampling rate of 8192~Hz. During measurement of the EEG, stimuli were presented using a laptop with the APEX 4 platform, developed at ExpORL \cite{francart_apex_2008} in conjunction with an RME Multiface II sound card and electromagnetically shielded Etymotic ER-3A insert phones. Experiments were conducted in an electromagnetically shielded and soundproofed cabin.

Preprocessing of stimuli and EEG recordings was performed in MATLAB. First, the EEG signal was downsampled from 8192~Hz to 1024~Hz, and artifacts were removed using a multichannel Wiener filter \cite{somers_generic_2018}. Then the EEG signal was re-referenced to a common average. All stimuli had an initial sampling frequency of 48~kHz. First, the envelope was estimated for all stimuli with a gammatone filterbank \cite{sondergaard_linear_2012, sondergaard_auditory_2013} with 28 subbands. Each subband envelope was estimated by taking the absolute value of each sample and raising it to the power of 0.6. All subbands were averaged to obtain 1 speech envelope \cite{biesmans_auditory-inspired_2017}. Finally, both EEG and stimuli envelopes were bandpass filtered between 0.5 and 32~Hz using a Chebyshev2 filter with an 80dB stopband attenuation and downsampled to 64 Hz.

For the Fairytale/held-out dataset, each recording was split into a train, validation, and test set containing 80\%, 10\%, and 10\% of each recording for each subject. The validation and test set were extracted from the middle of every recording to avoid unwanted effects at the edges of the recording (e.g., a subject not yet paying full attention or being startled). The remaining 80\% of the recording was added to the train set. None of the sets overlap so that after training the test set remains unseen for the model. Each recording was normalized separately by computing the mean and standard deviation for each EEG channel and the stimulus envelope on the train set. The mean was subtracted from the train, validation and test set and divided by the previously computed standard deviation.

The data of the MatrixEEG dataset was treated as a (single) test set, and each recording was normalized by subtracting the mean from each channel and stimulus envelope and dividing by the standard deviation (per channel). The Milan story was divided into a train/validation/test set and normalized in the same way as the recordings of the Fairytales dataset.

In the match/mismatch paradigm, all models are presented with 3 inputs: A segment of the EEG recording, the matching stimulus envelope segment and a mismatching (imposter) speech envelope segment. The imposter was extracted as specified by Monesi et al. \cite{monesi_lstm_2020}, i.e. 1 second after the matched stimulus envelope segment. If no imposter frame could be extracted (i.e. at the end of the recording), the segment was discarded from the dataset. Overlapping windows with 90\% overlap were extracted from each recording.

\subsection{Baseline models}
To compare the performance of our new model, we constructed 2 baseline models. The first baseline model is based on a state-of-the-art linear decoder. The second baseline model is based on CCA \cite{10.1088/1741-2552/abf771}, which obtains state-of-the-art performance when applied in a subject-specific setting.

\subsubsection{Decoder baseline}
\begin{figure*}[tb]
	
	\begin{tabular}[b]{c}
      \includegraphics[width=0.35\textwidth]{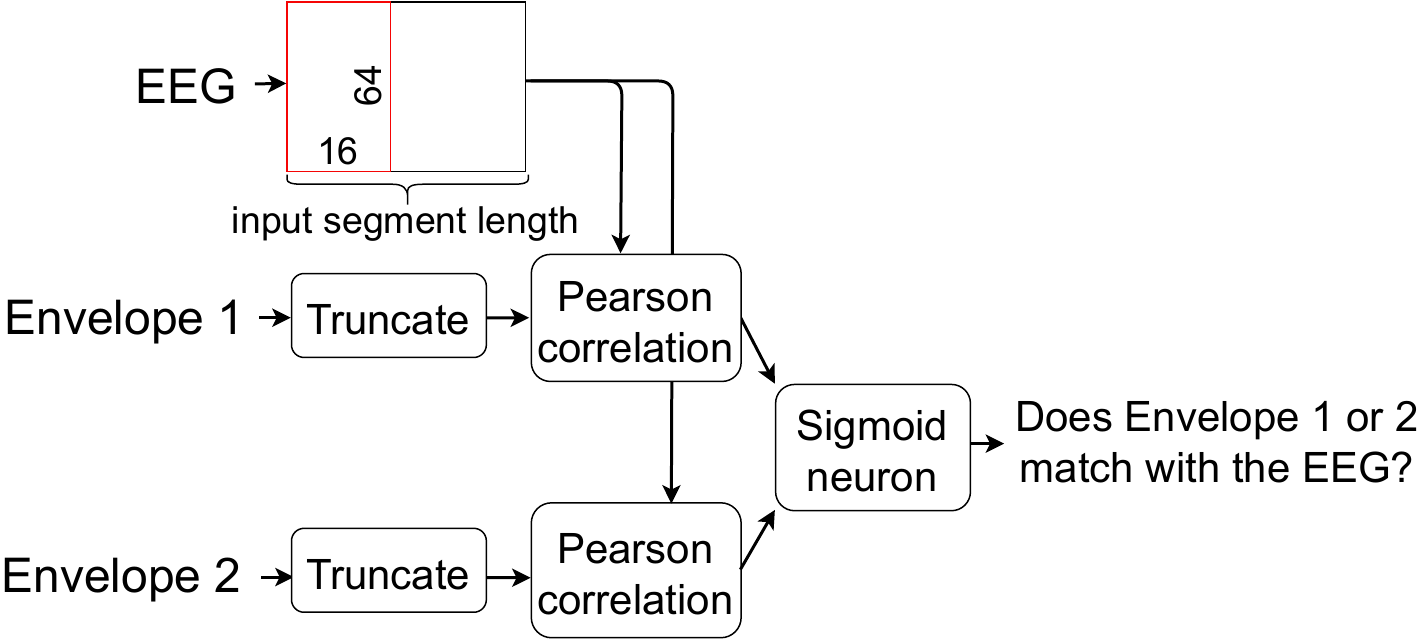}\label{fig:convolutional_baseline}\\
      \small (a) Baseline based on the linear decoder
    \end{tabular}
    \begin{tabular}[b]{c}
      \includegraphics[width=0.63\textwidth]{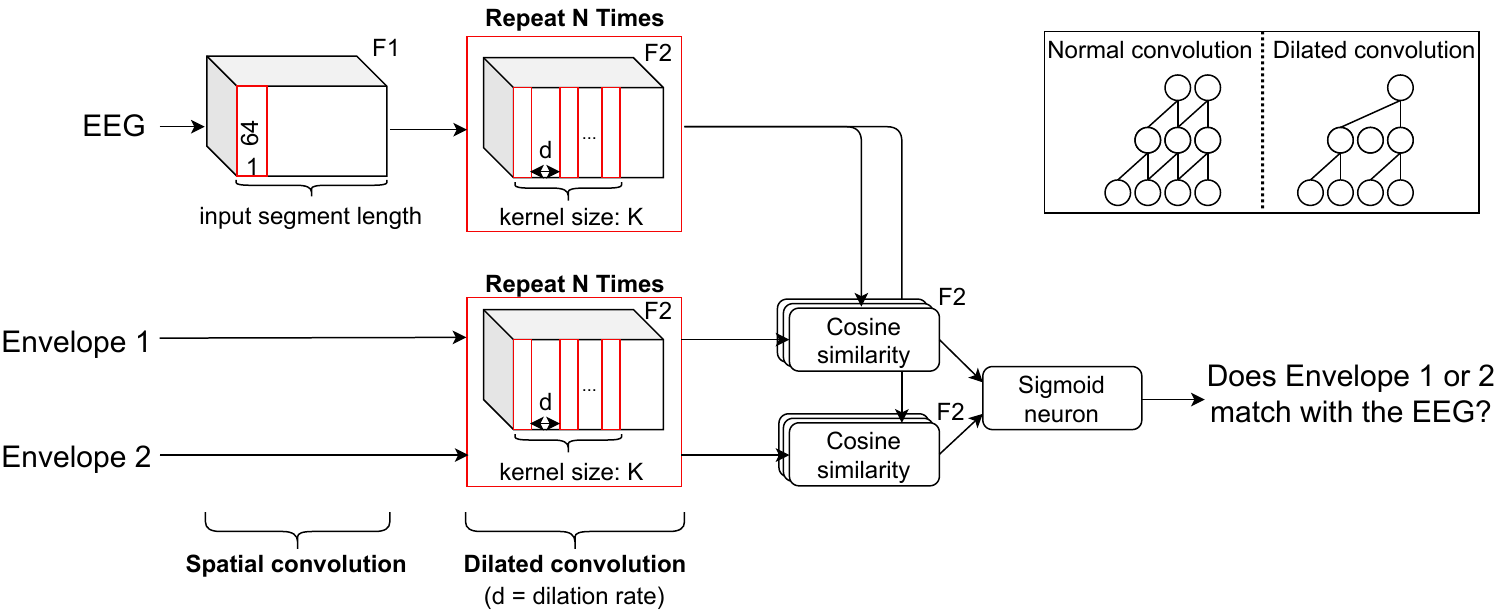}\label{fig:dilation_structure}\\
      \small (b) Dilated convolutional network \cite{accou_modeling_2020}.
    \end{tabular}
    
    \caption{The structure of the decoder baseline and dilated convolutional model}
	\label{fig:structure}
\end{figure*}

The decoder baseline is based on a linear decoder \cite{crosse_multivariate_2016, vanthornhout_speech_2018}, but adapted to be trained and evaluated in a match/mismatch paradigm. The integration window of the linear decoder is implemented as a convolution over the time dimension and across all EEG channels. As the EEG segment and envelope segments are time-aligned, the kernel of the convolution is functionally equivalent to the integration window of a linear decoder. After applying the convolution operation, a reconstruction of the stimulus envelope is obtained. This reconstructed envelope is compared to both envelope inputs with Pearson correlation. As the EEG and envelope input segments have the same length, it is impossible to reconstruct the last envelope samples because no EEG response is present in the selected EEG frame. Therefore, the envelope segments are truncated to the length of the reconstructed envelope segment. After correlating the reconstructed and presented envelope segments, the correlation coefficients are combined using a  sigmoid neuron to make a binary prediction (i.e. does envelope segment 1 or 2 match with the EEG segment?). A visual representation of this model can be seen in Figure \ref{fig:structure} (a).
The kernel size of the convolution was chosen to correspond to the integration window length that produces the highest correlation in linear decoders \cite{vanthornhout_speech_2018}, which is from 0-250ms. The decoder baseline model was implemented in Tensorflow (version 2.3.0) \cite{martinabadi_and_tensorflow_2015} with Keras \cite{chollet_francois_keras_2015} module.

\subsubsection{CCA}
Currently, CCA achieves state-of-the-art error rates for relating EEG with a speech stimulus representation in a subject-specific match/mismatch paradigm \cite{de_cheveigne_decoding_2018, 10.1088/1741-2552/abf771}. De Cheveigné et al.  \cite{10.1088/1741-2552/abf771} propose a gold-standard model (model G), to compare future work with. The baseline CCA model that we propose differs in a five ways from the proposed model G. Firstly, trials were selected across subjects, as this baseline model should be a subject-independent model to enable fair comparison with our proposed subject-independent dilated convolutional model. Secondly, no cross-validation is applied, as this is computationally infeasible due to the large amount of trials across subjects. Thirdly, we use different preprocessing (see section \ref{sec:prep} for more details about the specific preprocessing used in this paper). Fourthly, no PCA is applied to the EEG to reduce the number of channels to 32. In our subject-independent experiments, applying PCA to the EEG (per subject or across subjects) either decreased or did not significantly impact performance. Finally, the CCA model is evaluated using 2 different methods. In the first method, the classification is based on the difference of the distance between the transformed EEG and matched segment and the average distance between the transformed EEG and mismatched envelope segment, across trials per subject (as proposed by de Cheveigné et al. \cite{10.1088/1741-2552/abf771}. In the second method, a single imposter segment is selected 1 second after the end of a matching segment (see Figure \ref{fig:matchmismatch}), following the same procedure as used for the dilated convolutional model and the decoder baseline model (see section \ref{sec:prep} for more information). These models will be annotated as "CCA (averaged)" and "CCA" in all figures, respectively. According to de Cheveigné et al~\cite{10.1088/1741-2552/abf771}, using an aggregate of mismatched windows is more robust and increases correlations and classification performance. However, this approach is also not straightforward to implement for the dilated convolutional model. For completeness, both versions are included for CCA and compared to the dilated convolutional model. All CCA models were implemented using the NoiseTools package (version 12-May-2021) \cite{de_cheveigne_alain_noisetools_nodate} in MATLAB 2020b.

\subsection{Dilated convolutional model}
The dilated convolutional model consists of 4 steps. In the first step, the EEG channels are linearly combined from 64 to $F_1$ using a 1D convolution with a kernel size of 1 and a filter size of $F_1$. Then, $N$ repeated dilated convolutional layers using $F_2$ filters with a kernel size of $K$ are applied to both EEG and envelope segments. After each dilated convolutional layer, a rectified linear unit (ReLU) non-linearity \cite{nair_rectified_2010} is applied. The weights of the dilated convolutional layers for the envelopes are shared for both envelope inputs. After non-linearly transforming the EEG and envelope data, the EEG representations are compared to both envelopes using cosine similarity. Finally, these similarity scores are fed into a single neuron with a sigmoid non-linearity to generate a prediction. After a hyperparameter sweep, values 8 and 16 were chosen for $F_1$ and $F_2$, respectively. The dilated convolutional model was implemented in Tensorflow (version 2.3.0) \cite{martinabadi_and_tensorflow_2015} with Keras \cite{chollet_francois_keras_2015} module.

\subsection{Model training}
 Both the decoder baseline and the dilated convolutional model are implemented in used an Adam optimizer with a learning rate of 0.001 and binary cross-entropy as a loss function. Models were trained for maximally 400 epochs, and early stopping was used based on the validation loss with a patience factor of 5 epochs. If the training was stopped early, the model's weights were restored to their value in the epoch with the lowest validation loss. 
 The CCA models were trained following the procedure for model G in de Cheveigné et al. \cite{10.1088/1741-2552/abf771}.
 All experiments used input segment lengths of 10 seconds unless indicated otherwise. All models were trained in a subject-independent way, i.e., they received data from multiple subjects during training on the train set of the Fairytales dataset.

\section{Evaluation of the dilated convolutional model}
\label{sec:mm_experiments}
To showcase the performance of the dilated convolutional model, the influence of segment length, frequency range and receptive field size were evaluated in this section. From an application perspective, it is important to know whether the model generalizes well to unseen data. If this is not the case, extra subject-specific fine-tuning might be necessary if the model's performance is not high enough for clinical purposes. Therefore, fine-tuning on unseen subjects was also evaluated.

\subsection{Input segment length}
\label{sec:input_segment_length}
In auditory attention decoding, increasing the length of input segments increases the model performance, as the model receives more data for making a single prediction \cite{deckers_eeg-based_2018}. As this also applies to the match/mismatch paradigm for CCA \cite{10.1088/1741-2552/abf771, de_cheveigne_decoding_2018}, the performance of the dilated convolutional model and the baseline models should increase with longer input segments.

\subsubsection{Setup}
The dilated convolutional and decoder baseline were trained and evaluated on the Fairytale dataset for input segment lengths of 0.5, 1, 2, 5, 10 and 20 seconds. The CCA models were trained using the full train set of the Fairytale data. The results for all models were compared for each input segment length and statistically evaluated utilizing a linear mixed-effects model, using the \verb|lme4| (version 1.1.23) \cite{bates_lme4_2021} and \verb|lmerTest| \cite{kuznetsova_lmertest_2020} (version 3.1.2) packages in R (version 4.0.3)\cite{Team2014RAL}. Input segment length and model were designated to be fixed effects and the subject was designated as a random effect.The \verb|emmeans| (1.4.6) \cite{lenth_emmeans_2021} package was used to perform a pairwise Tukey's test on the levels of the model fixed effect to determine whether the performance of between models differed significantly.

\subsubsection{Results}
\begin{figure}
    \centering
    \includegraphics[width=0.49\textwidth]{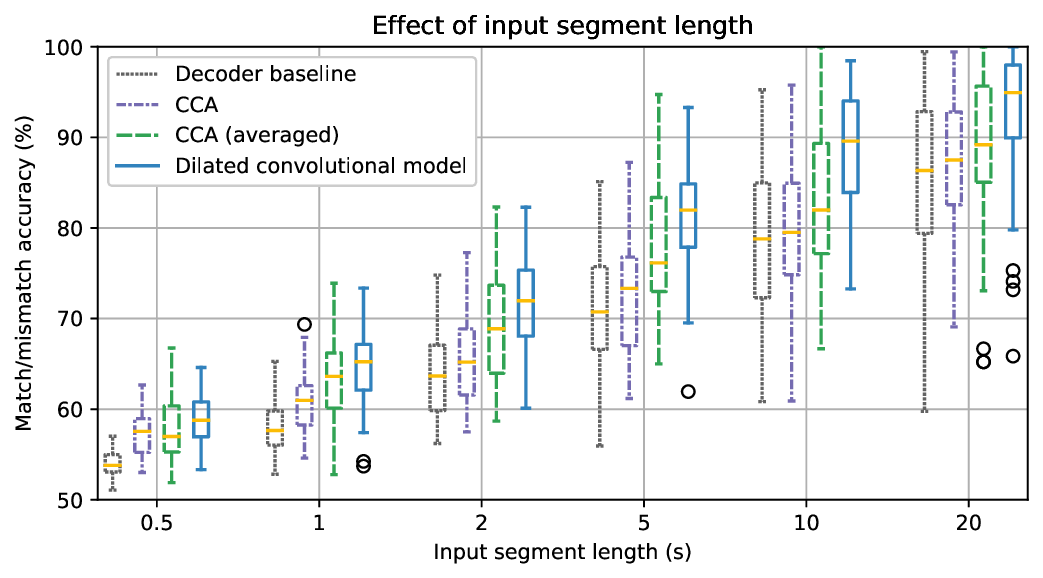}
    \caption{Each point in the boxplot is the accuracy for one subject averaged across recordings on the (unseen) test set. Hollow circles represent outliers. Performance increases with larger input segment lengths (p $\leq 10^{-15}$), as does variance for the decoder baseline model. The dilated model had a receptive field size of 27 samples ($\simeq$ 420 ms). The dilated convolutional model significantly outperforms all baseline models (p $\leq 0.001$). }
    
    \label{fig:input_segment_length}
\end{figure}

The performance for the dilated convolutional model and all baseline models is displayed in Figure \ref{fig:input_segment_length}. The performance of all models increases with input segment length. Note that the variability in the decoder baseline model also increases with increasing input segment length (e.g., at 20 seconds, the accuracy ranged from 60\% to 100\%, revealing major inter-subject differences). The effect of both model type and input segment length were significant (p $\leq 10^{-4}$ and p $\leq 10^{-15}$ respectively). Further comparison within the model types with the Tukey's test revealed that the performance of all models was significantly different (p $\leq 0.001$)

\subsubsection{Discussion}
The increase in performance by enlarging the input segment length is probably due to the model having more data to decide on. The same trend can be seen in auditory attention detection \cite{deckers_eeg-based_2018, taillez_machine_2017} and in previous literature for CCA \cite{de_cheveigne_decoding_2018, 10.1088/1741-2552/abf771}. Caution should be advised when using large input segment lengths (e.g., 20 seconds) as ceiling effects may occur due to some subjects obtaining the maximal score on the test set, at which point no further improvement can be gained.

\subsection{Frequency band}
\label{sec:freq_band}

EEG signals are usually evaluated in different frequency bands: $\delta$ (0.5-4Hz), $\theta$ (4-8Hz), $\alpha$ (8-14Hz) and $\beta$ (14-32Hz). As shown in previous literature \cite{vanthornhout_speech_2018, ding_emergence_2012}, linear decoders perform optimally in the $\delta$-band. De Cheveigné et al. \cite{de_cheveigne_decoding_2018} also show higher performance for lower frequencies. To evaluate whether this is also the case for our baseline models and the dilated convolutional model, all models were trained and evaluated on (combinations of) different frequency bands.

\subsubsection{Setup}
For this experiment, both EEG and stimulus envelopes of the Fairytales dataset were bandpass filtered with a Chebyshev2 filter (order of 2000, 80dB stopband attenuation, 1dB passband ripple) for all possible bands and combinations of bands ($\delta + \theta$, $\delta + \theta + \alpha$, $\delta + \theta + \alpha + \beta$) instead of the previously stated 0.5-32~Hz in Section \ref{sec:prep}. Both the dilated convolutional and baseline models were trained and evaluated on the resulting data for each (combination of) frequency band(s). Finally, all models were evaluated using a Wilcoxon signed-rank test with Holm-Bonferroni correction for each band (combination) separately.

\subsubsection{Results}
\begin{figure}
    \centering
    \includegraphics[width=0.49\textwidth]{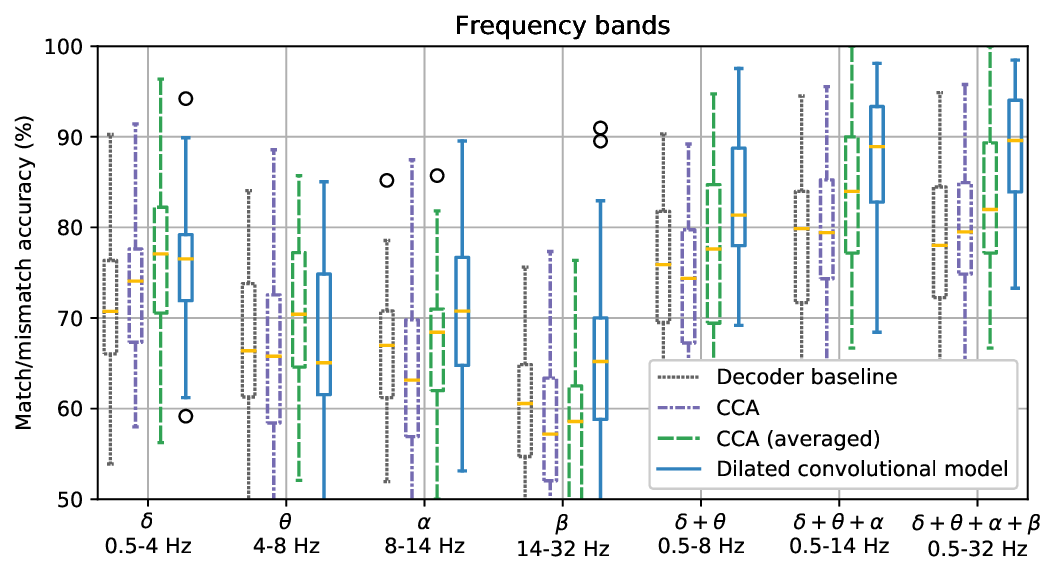}
    \caption{Each point in the boxplot is the accuracy for one subject averaged across recordings on the (unseen) test set. Hollow circles represent outliers. The dilated model had a receptive field size of 27 samples ($\simeq$ 420 ms). The baselines and dilated convolutional model are all trained and evaluated on different frequency bands. Combining multiple frequency bands increases performance for all models. For individual bands, higher performance is obtained for lower frequency bands.}
    \label{fig:freq_bands}
\end{figure}

The performance for all models increased by adding more frequency components (as can be seen in Figure \ref{fig:freq_bands}). Looking at single frequency bands, all models have decreased performance for higher frequency bands. Each model performed significantly different (p $\leq$ 0.050) for the combined bands, with the dilated convolutional model outperforming all the baseline models. In the $\delta$ band, the dilated convolutional model and CCA (averaged) both significantly outperformed the decoder baseline and CCA model (p $\leq$ 0.010), but no significant difference in performance was found between the CCA (averaged) and the dilated convolutional model. In the $\theta$ band, the CCA (averaged) significantly outperformed the CCA model (p $\leq$ 0.001). The performance of all models differed significantly in the $\alpha$ band (p $\leq$ 0.050), except the CCA (averaged) and decoder baseline model. Finally, all models performed significantly differently in the $\beta$ band (p $\leq$ 0.010), except CCA and CCA (averaged), and CCA and the decoder baseline.

\subsubsection{Discussion}
When looking at single frequency bands, all models perform better in lower frequency bands. Crosse et al.\cite{Crosse14195} have shown the same trend for linear decoders, de Cheveigné et al.\cite{de_cheveigne_decoding_2018} have shown this for CCA . The best performance for single frequency bands is obtained in the $\delta$-band.  In literature, it has been shown that individual speech recognition is linked to the responses in the $\delta$-band \cite{ding_emergence_2012, ding_cortical_2014,doelling_acoustic_2014, woodfield_role_2010, vanthornhout_speech_2018, lesenfants_predicting_2019}.

When combining multiple frequency bands, all models perform better, suggesting they can leverage additional information from multiple frequency bands.

\subsection{Receptive field size}
\label{sec:receptive_field_size}
In linear decoders, an integration window is used to compensate for the delayed brain response. Optimally, integration windows of around 250~ms are used~\cite{vanthornhout_speech_2018, osullivan_attentional_2015}. For CCA, time-lags can be applied to both the EEG and envelope to compensate for this \cite{de_cheveigne_decoding_2018, 10.1088/1741-2552/abf771}, as well as a relative shift where EEG is advanced with regards to the envelope \cite{10.1088/1741-2552/abf771}. Model G in \cite{10.1088/1741-2552/abf771} and our CCA baselines advance the EEG relative with 200~ms, and apply timelags from 0-250~ms for both EEG and envelopes . In the dilated convolutional model, there is no explicit integration window. However, as shown in Oord et al. \cite{oord_wavenet:_2016}, we can define the receptive field size of the dilated convolutions as the number of input samples involved in computing a single output sample. This receptive field size can be modified in the dilated convolutional model by varying the kernel size and the number of consecutive dilated convolutions. The receptive field size is equal to $K^{N}$, where $N$ denotes the number of layers and $K$ is the kernel size in samples.

\subsubsection{Setup}
A parameter sweep was executed for input segments of 10 seconds to determine which receptive field size yields optimal results. For kernel sizes 2, 3 and 4, all possible depths were explored (i.e. until the receptive field size became larger than the input segment). A Wilcoxon signed ranked test with Holm-Bonferroni correction was used to compare each dilated convolutional model with the baseline models.

\subsubsection{Results}
\begin{figure*}
    \centering
    \includegraphics[width=\textwidth]{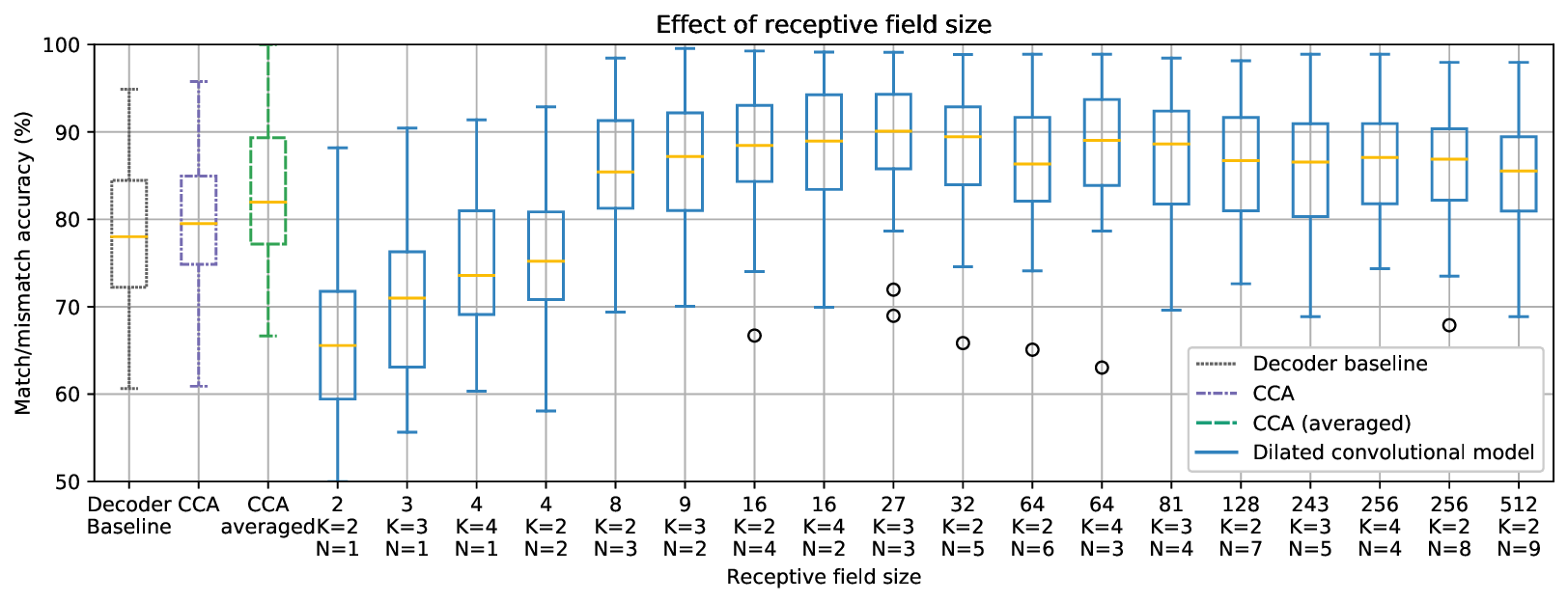}
    \caption{Each point in the boxplot is the accuracy for one subject averaged across recordings on the (unseen) test set.  Hollow circles represent outliers. The dilated convolutional model is trained for different receptive field sizes. This is done by varying the number of consecutive dilated convolutional layers ($N$) and kernel size ($K$). The optimal performance is reached for 27 samples, which corresponds to $\sim$ 420~ms.}
    \label{fig:receptive_field}
\end{figure*}

The results for different receptive field sizes are displayed in Figure \ref{fig:receptive_field}. The performance rises with increasing receptive field size until 27 samples ($\approx$ 420~ms), after which performance declines. Compared to the baseline models, the dilated convolutional model significantly outperforms all baseline models when using a receptive field size between 16 and 32 samples (250~ms to 500~ms) (p $\leq$ 0.001). Additionally, the dilated convolutional model significantly outperforms the CCA (averaged) model for configurations $K=4,N=3$ (64 samples, 1 second) and $K=3,N=4$ (81 samples, 1.27 seconds)(p $\leq 10^{-8}$). The dilated convolutional model also outperforms the CCA and decoder baseline model significantly when using a receptive field size between 8 and 512 (125~ms and 8 seconds)(p $\leq 10^{-4}$). Both CCA models outperform the dilated convolutional model significantly for receptive field sizes $\leq$ 4 ($\leq$ 63~ms)(p $\leq$ 0.001), while the decoder baseline significantly outperforms the dilated convolutional model for receptive field sizes $\leq$ 3 ($\leq$ 47~ms)(p $\leq 10^{-7}$).

\subsubsection{Discussion}
The dilated convolutional model performs best with a receptive field size of 27 samples, which corresponds to 420~ms. This is very long compared to the best performing integration window of the linear decoder, which is from 0 to 250~ms \cite{vanthornhout_speech_2018, osullivan_attentional_2015}, or the timelags introduced for CCA (250~ms)\cite{10.1088/1741-2552/abf771}. The need for this long receptive field size might be explained by the dilated convolutional network's non-linear nature and bigger size \cite{diliberto_low-frequency_2015}. Due to this, the model might capture later responses more effectively.

\subsection{Generalization}
\label{sec:generalization}

The test set of the Fairytales dataset contains data extracted from the middle of individual recordings, which all models have not seen during training. However, as the training data is extracted from the same recording, the models have seen the \emph{subjects} from the test set during training. Therefore, it is important to check whether the dilated convolutional model still performs well on subjects not present in the train set for our application perspective.

\subsubsection{Setup}
The dilated convolutional model was trained on the Fairytales dataset and evaluated on the test set of the 20 remaining (i.e., unseen) subjects of the Fairytales held-out set. To test if there was a significant difference between the performance, the Mann-Whitney U-test was used. From an application perspective, it is also useful to know how many subjects are necessary to saturate the generalizability of the dilated convolutional model, i.e. how many subjects are needed in training to ensure good performance on unseen subjects. For this experiment, the dilated convolutional model was trained on a varying number of subjects (1-48) from the Fairytales dataset training set and evaluated on the held-out Fairytales test set.

\subsubsection{Results}
\begin{figure}
    \centering
    \includegraphics[width=0.49\textwidth]{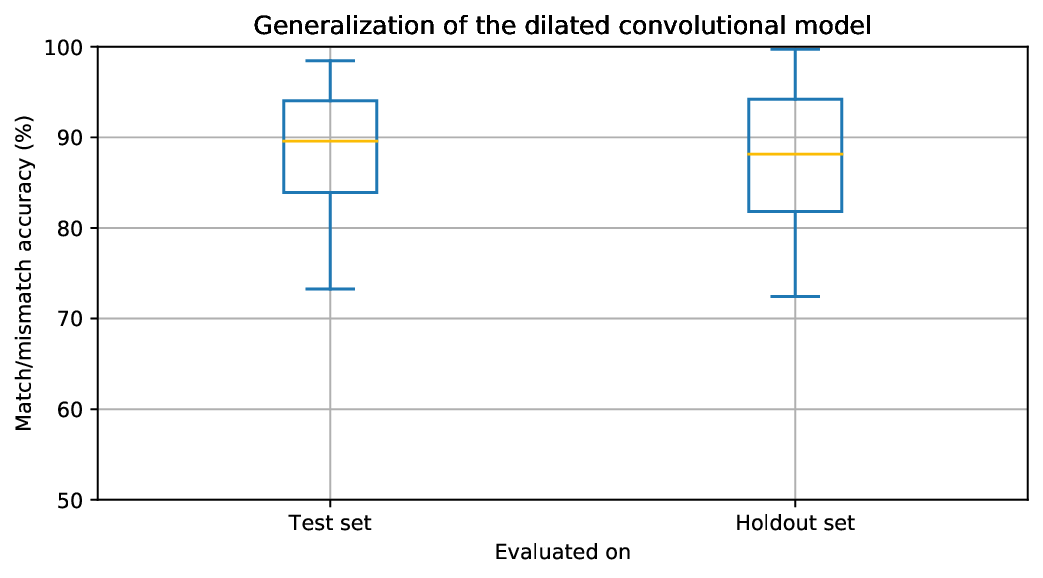}
    \caption{Each point in the boxplot is the accuracy for one subject averaged across recordings. A dilated convolutional model trained on the Fairytales dataset is evaluated on the test sets of the Fairytales dataset and the held-out dataset. The dilated model had a receptive field size of 27 samples ($\simeq$ 420 ms). The performance for the Fairytales test set, containing subjects that the model has seen during training, does not significantly differ from the performance held-out dataset, containing only unseen subjects.}
    \label{fig:held-out}
\end{figure}

\begin{figure*}
    \centering
    \includegraphics[width=\textwidth]{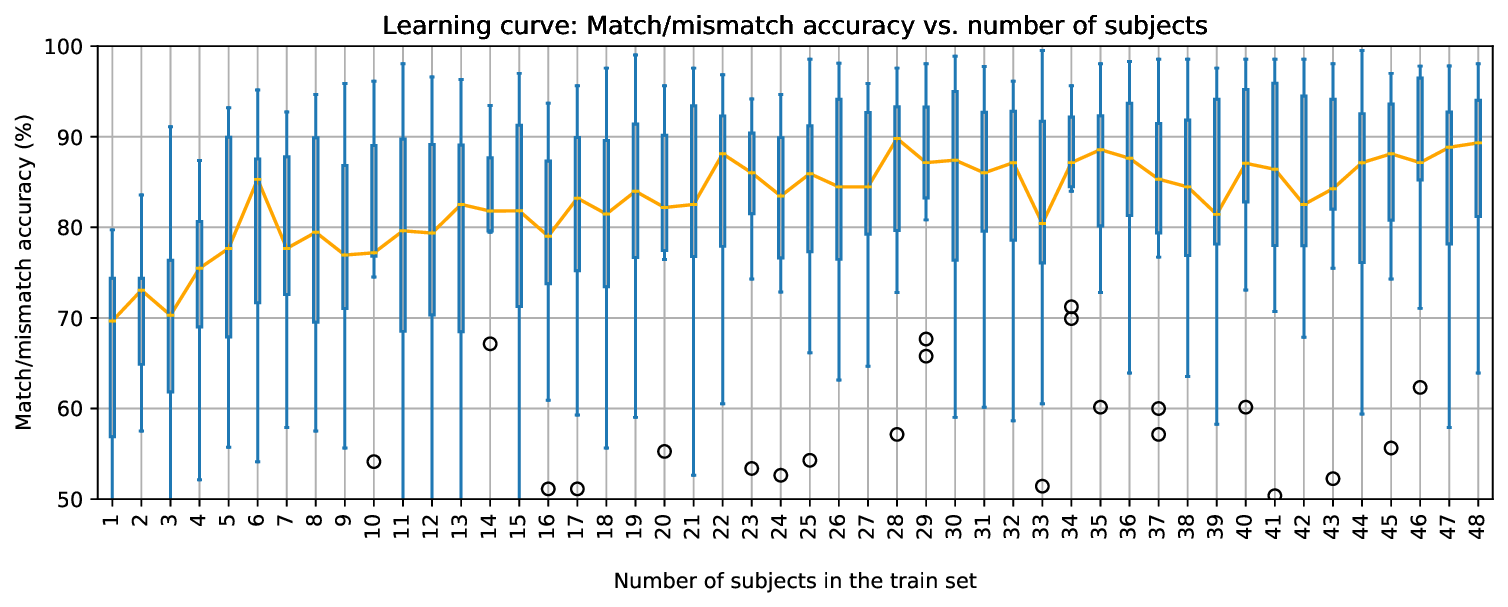}
    \caption{Learning curve. Each point in the boxplot is the accuracy for one subject averaged across recordings.  Hollow circles represent outliers. The dilated convolutional model is trained on a variable number of subjects in the Fairytales dataset and evaluated on the (unseen) held-out Fairytales test set. The orange line connects the median accuracies obtained on the held-out dataset. Including additional subjects in training increases performance until optimal performance is reached (28 subjects). After this point, performance still fluctuates but does not increase anymore.}
    \label{fig:learning_curve}
\end{figure*}

In Figure \ref{fig:held-out}, the performance of the dilated convolutional model for the Fairytales test set and the held-out set are shown. The difference between the scores for both test sets is not significantly different (W=449, p=0.3407). The learning curve in Figure \ref{fig:learning_curve} indicates that the generalization saturates at approximately 28 subjects.

\subsubsection{Discussion}
The learning curve (Figure \ref{fig:learning_curve}) and held-out dataset performance (Figure \ref{fig:held-out}) confirm that the dilated convolutional model generalizes well to unseen subjects. This is especially interesting from an application perspective, as it removes the need to collect training data from a prospective subject. This can substantially reduce the time spent during hearing assessment procedures. Only 28 subjects are needed to train a well-performing generalizable model.

\subsection{Fine-tuning}
\label{sec:fine-tuning}

A plausible way to increase performance is to fine-tune a pre-trained model to an unseen subject. By doing this, the model is effectively transformed from a subject-independent model to a subject-specific model for the new subject.

\subsubsection{Setup}
The dilated convolutional model was trained on the Fairytales dataset and fine-tuned on the training set of each subject of the held-out dataset separately. Then, the fine-tuned models were compared to the performance of the subject-independent model on the held-out dataset with a Wilcoxon signed-rank test. A learning curve was constructed to see how much data was needed for each subject to reach equilibrium. As weights can also be frozen and only specific layers can be tuned, an experiment was also conducted where different layers of the dilated convolutional model were grouped (spatial EEG layer, dilated convolutional EEG layers, dilated convolutional envelope layers and the output layer) and the permutation of each grouping was evaluated. To compare the performance to the baseline, a Wilcoxon signed-rank test was used with Holm-Bonferroni correction.

\subsubsection{Results}
Fine-tuning the pre-trained dilated convolutional model on each subject in the held-out set separately improved performance for all but 1 subject. The increase in performance overall is statistically significant (W=3, p=0.001). The learning curve in Figure  \ref{fig:fine_tune_learning_curve} shows increasing performance on the held-out dataset up until 1 hour of data for each subject. In Figure \ref{fig:fine_tune_layers}, the results of fine-tuning different groups of layers are shown. Each configuration significantly outperforms the dilated convolutional model without retraining (p $\leq$ 0.050).

\subsubsection{Discussion}
\label{sec:disc:fine_tuning}
\begin{figure}
    \centering
    \includegraphics[width=0.49\textwidth]{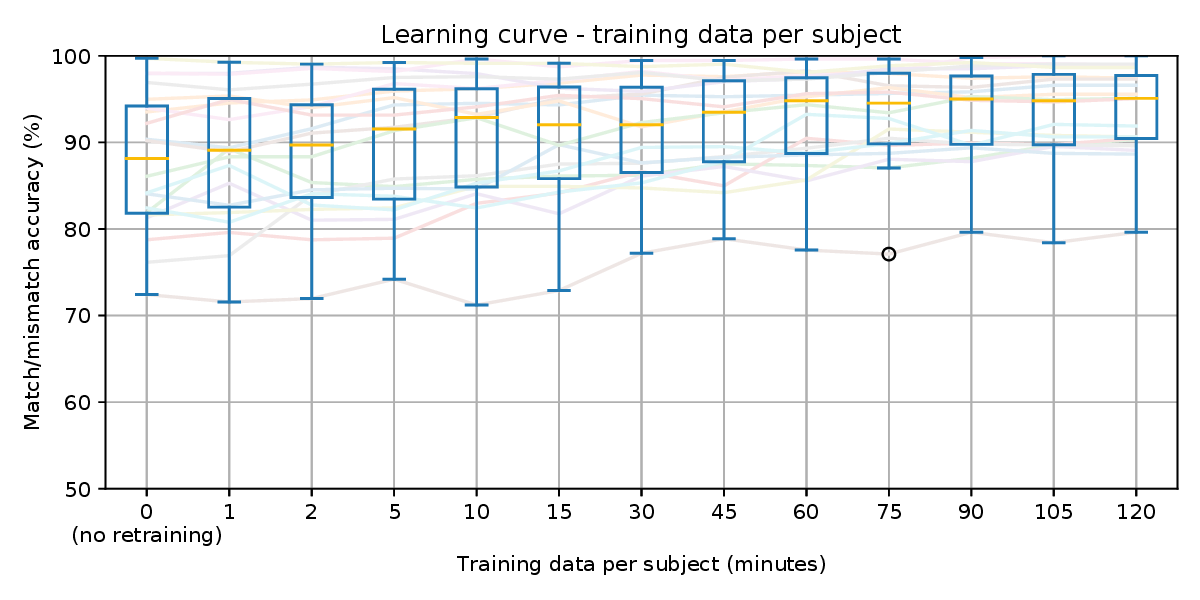}
    \caption{Each point in the boxplot is the accuracy for one subject averaged across recordings. Hollow circles represent outliers. A dilated convolutional model was fine-tuned on the train set of the held-out dataset with varying amounts of training data per subject, as displayed on the x-axis. The dilated model had a receptive field size of 27 samples ($\simeq$ 420 ms). An input segment length of 10 seconds was used. The model is then evaluated on the (unseen) test set of the held-out dataset for each subject separately. Including more training data for each subject increases performance up until 60 minutes, after which performance stagnates.}
    \label{fig:fine_tune_learning_curve}
\end{figure}

\begin{figure*}
    \centering
    \includegraphics[width=\textwidth]{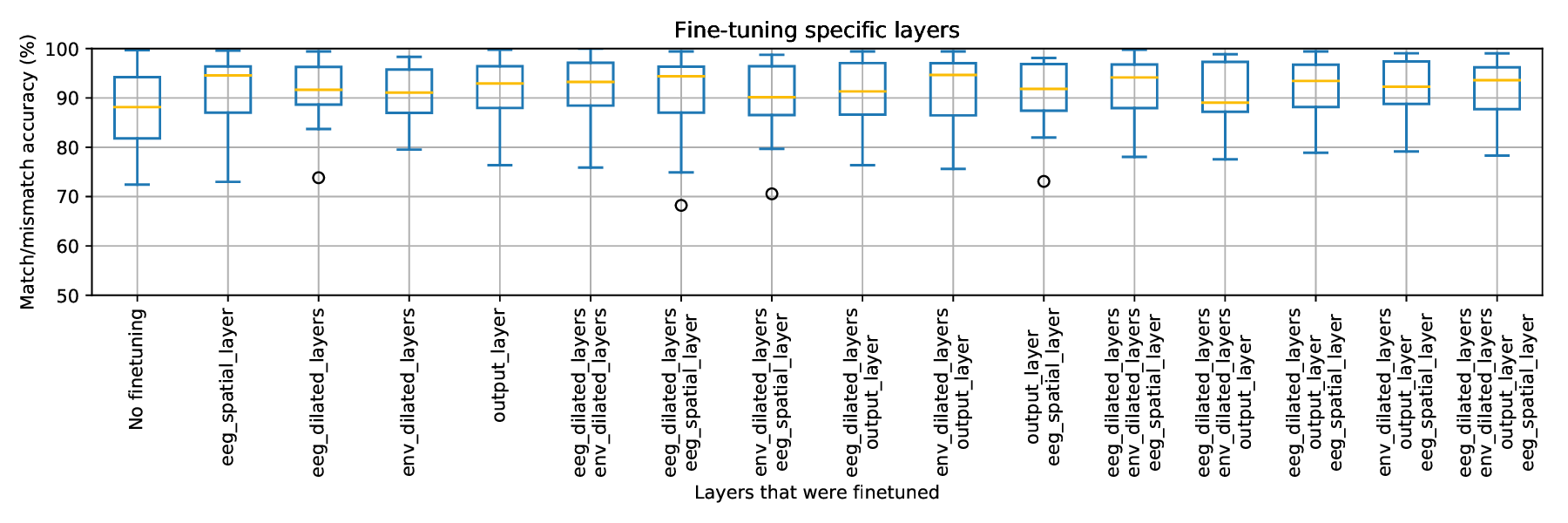}
    \caption{Each point in the boxplot is the accuracy for one subject averaged across recordings. Hollow circles represent outliers.  The dilated model had a receptive field size of 27 samples ($\simeq$ 420 ms). The dilated convolutional model is fine-tuned on the training set of the held-out dataset and evaluated on the (unseen) test set of the held-out test set with different layers frozen. On the x-axis, the fine-tuned layers are listed, omitted layers were frozen during fine-tuning.}
    \label{fig:fine_tune_layers}
\end{figure*}

Fine-tuning on the held-out training set has shown that performance can significantly increase for most subjects by fine-tuning the model on a specific subject. This boost in performance can be used in Section \ref{sec:beh_vs_obj} to improve sigmoid fits, which might allow SRT estimation for subjects for whom the fits failed previously. In practice, this would come at the cost of collecting a small amount of training data from the new subject, which would increase the time duration of data collection. Looking at Figure \ref{fig:fine_tune_learning_curve}, the model's performance seems to saturate when using more than 60 minutes of data. As shown in Figure \ref{fig:fine_tune_layers}, every fine-tuning scheme will result in significantly higher performance.

\section{Speech intelligibility prediction}
\label{sec:beh_experiments}
We established a link between the speech reception threshold (SRT) of the (behavioral) MATRIX test in Flemish \cite{luts_development_2014} and the performance of the dilated convolutional model, to investigate whether our model performance can be used as a proxy of speech intelligibility. Like the MATRIX test, we define speech intelligibility as word recognition accuracy in noise.

\subsection{Comparison to state-of-the-art behavioral testing}
\label{sec:beh_vs_obj}

\subsubsection{Setup}
To evaluate the dilated convolutional model as an objective measure of speech intelligibility, a dilated convolutional model was trained on the  Fairytales dataset with an input segment length of 20 seconds and evaluated on the MatrixEEG dataset. The model was evaluated per subject for each SNR separately. Per subject, the relationship between SNR and accuracy was modeled using a psychometric curve (see Equation \ref{eqn:pyscho}) \cite{vanthornhout_speech_2018}.

\begin{equation}
    \label{eqn:pyscho}
    Accuracy(SNR) = \gamma + (1 - \gamma -\lambda) * \frac{1}{1 + \exp{-\frac{SNR-\alpha}{\beta}}}
\end{equation}

The accuracy for the condition without noise was discarded before fitting, the guess rate ($\gamma$) and lapse rate ($\lambda$) were fixed to 0.5 and 0, respectively, while the boundary values for the slope ($\beta$) were set to 0.05 and 50. There were no boundary conditions for the midpoint ($\alpha$). These boundary values were similar to the ones used in Vanthornhout et al \cite{vanthornhout_speech_2018}. Fitting was performed using the non-linear least-squares implementation of SciPy \cite{virtanen_scipy_2020} (\verb|scipy.optimize.curve_fit|). Fits coinciding with the boundary conditions were discarded. The midpoints of these newly fitted sigmoids were correlated, using a Pearson correlation, with the SRT as found by the behavioral MATRIX test. In this way, a link between the accuracy of dilated convolutional and speech intelligibility as estimated by a behavioral test can be established.
The Milan fairytale data (see \ref{sec:matrix_eeg}) can be used to finetune the model, gaining overall better performance and improving sigmoid fits. However, seeing that data collection in clinical practice is difficult and costly, we will restrict finetuning to the subjects for which the sigmoid fitting failed.

To estimate the influence of the 0.5dB test-rest variability of the behavioral MATRIX test \cite{luts_development_2014}, we used a Monte-Carlo simulation of 100000 iterations. In this simulation, random Gaussian noise with a mean of 0 and a standard deviation of 0.5~dB was added to the SRT values of the behavioral MATRIX test, before the Pearson correlation between the SRT's of the behavioral MATRIX test and midpoints of the sigmoids was calculated.

\subsubsection{Results}
\begin{figure}
    \centering
    \includegraphics[width=0.49\textwidth]{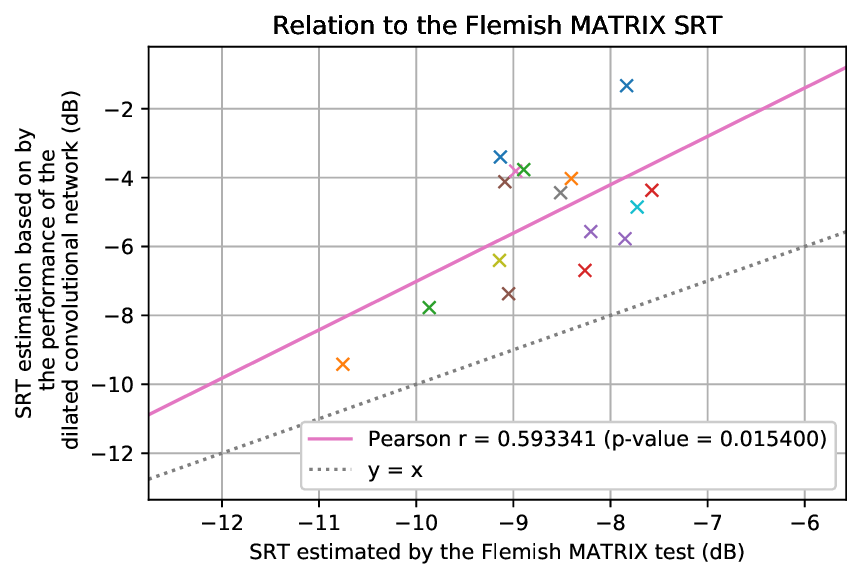}
    \caption{Comparison between the midpoints of the sigmoids fitted on the performance of the dilated convolutional model and the behavioral MATRIX score for the MatrixEEG dataset. Each cross corresponds to a single subject. 4 subjects were excluded due to the bad fit of the sigmoid. A significant relationship is found between the midpoints of the fitted sigmoids and the SRT as estimated by the behavioral MATRIX test.}
    \label{fig:obj_vs_beh}
\end{figure}

\begin{figure}
    \centering
    \includegraphics[width=0.49\textwidth]{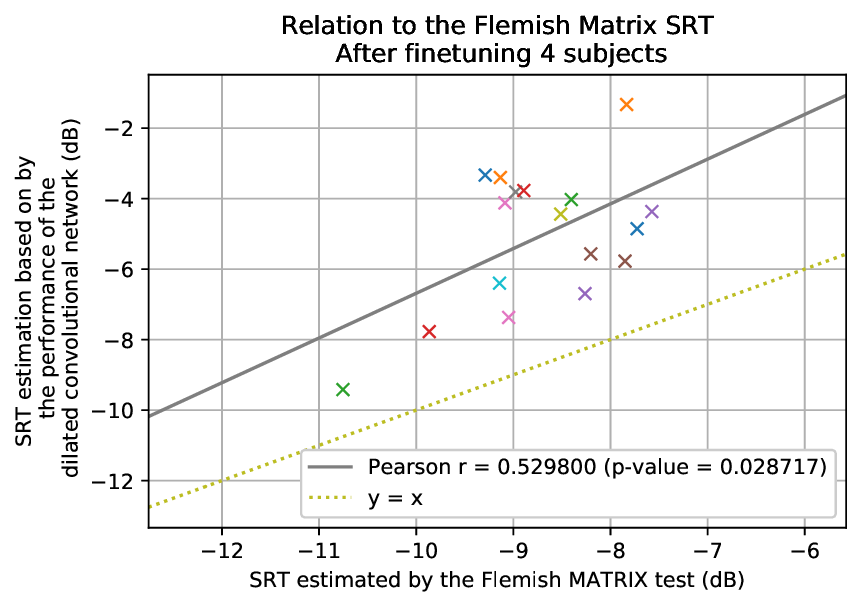}
    \caption{Comparison between the midpoints of the sigmoids fitted on the performance of the dilated convolutional model and the behavioral MATRIX score for the MatrixEEG dataset. Each cross corresponds to a single subject. Sigmoid fitting improved for one out of the 4 subjects that had bad fitting previously in Figure \ref{fig:obj_vs_beh} by fine-tuning the dilated convolutional model to these subjects. The relationship between the midpoints of the fitted sigmoids and the SRT as estimated by the behavioral MATRIX test is still significant after adding this subject, although it slightly decreased (0.59 vs. 0.52).}
    \label{fig:obj_vs_beh_finetuned}
\end{figure}

Fits for 4 sigmoids reached boundary conditions and were discarded from further analysis. The remaining 16 midpoints of the sigmoids are significantly correlated with the SRT's as found with the MATRIX test (r=0.59, p=0.0154), as seen in Figure \ref{fig:obj_vs_beh}. Finetuning on the Milan fairytale data improved one of 4 sigmoid fits that previously failed. Adding this subject to the analysis decreased the overall correlation to 0.53 (p=0.0287), as can be seen in Figure \ref{fig:obj_vs_beh_finetuned}.

The Monte-Carlo simulations yielded 0.5129 (95\% confidence interval = [0.5122, 0.5135]) and 0.4556 (95\% confidence interval = [0.4549, 4563]) as average Pearson correlation value across iterations for the non-finetuned and the finetuned case respectively.

\subsubsection{Discussion}
In Figure \ref{fig:obj_vs_beh}, a significant correlation is shown between the golden standard behavioral MATRIX test and the sigmoids fitted on the accuracies of the dilated convolutional model. This suggests that the dilated convolutional model can be used as an objective proxy or alternative for the MATRIX test. Similar results have been shown using subject-specific linear decoders \cite{vanthornhout_speech_2018}. While the correlation coefficient between objective measure and Matrix SRT is lower compared to Vanthornhout et al. \cite{vanthornhout_speech_2018} (0.59 vs. 0.69); our approach has multiple advantages. Firstly, a pre-trained subject-independent model is used, eliminating the need to collect training data for new subjects to evaluate. Secondly, if the sigmoid fitting fails, it is possible to improve the model by fine-tuning it to the specific subject, although at the cost of collecting training data for that subject. Lastly, the linear trend between the objective and behavioral SRT estimation (i.e. the SRT as estimated by the behavioral MATRIX test and the midpoints of the sigmoids fitted on the accuracy scores of the dilated convolutional model) seems to be parallel to $y=x$, which makes sense because the target of the behavioral measure (50\% correct word score) is arbitrary, leading to an arbitrary offset to our objective measure. It should be noted, however, that the performance of our objective method still could depend on a number of cognitive effects, such as attention \cite{osullivan_attentional_2015, ding_neural_2011}. 

Apart from the potential imprecision of the objective measure, the remaining differences between objective and subjective measures can be due to (1) imprecision of the behavioral measure (as our Monte-Carlo simulation reveals that the test-retest reliability can reduce the correlation with 0.08) (2) inherent differences between (presumably) decoding the acoustic representation of speech from the brain and engaging the entire auditory/language/memory circuits of the brain. 

To improve SRT estimation accuracy, the input segment length may be increased at the cost of needing more data. Another possibility is using a more flexible method than the sigmoid fitting to find a value to correlate with the MATRIX SRT, e.g., a simple artificial neural network.

\section{Discussion}
\label{sec:Discussion}

This paper evaluated both the robustness and viability of the dilated convolutional model as a proxy for speech intelligibility. Section \ref{sec:input_segment_length} and Section \ref{sec:freq_band} showed that model performance increases with longer input segment lengths and broader frequency bands. Furthermore, the dilated convolutional model benefits from a large receptive field compared to the integration windows of linear models (420~ms vs. 250~ms), as shown in Section,\ref{sec:receptive_field_size}. Furthermore, in a move towards better applicability, in Section \ref{sec:generalization} and Section \ref{sec:fine-tuning} generalisability to subjects unseen during training was evaluated, and the possibility to increase model performance through finetuning. Lastly, in Section \ref{sec:beh_vs_obj}, it was shown that the dilated convolutional model could be used as an effective proxy of speech intelligibility, even on completely unseen data (unseen subjects and speech material), which is the main benefit of a subject-independent model.

While model performance is promising, the non-linear nature of the model makes it hard to evaluate and interpret what feature of the data it is using to base predictions on. In literature, efforts have been made to create a locally linear version of non-linear models\cite{keshishian_estimating_2020}. The same techniques were used on the dilated convolutional model, but the results proved difficult to interpret and did not seem biologically plausible. Further work should aim to extract neuroscientific knowledge from the model and move away from the "black box" paradigm common for non-linear neural networks.

Currently, the speech envelope was used as the only speech feature input to the network. In future work, models may benefit from less coarse features such as a mel spectrogram\cite{akbari_towards_2019} or more advanced features such as phoneme identity \cite{diliberto_low-frequency_2015} or word embeddings\cite{broderick_electrophysiological_2018}, possibly in a hierarchical model.

\section{Conclusion}
This paper introduced a dilated convolutional neural network to model the relationship between EEG and acoustic stimulus, which can be trained in a subject-independent manner. We established a significant correlation between our model and the current golden standard in behavioral auditory testing, signifying that our model can be used as a proxy for speech intelligibility, even on previously unseen subjects. Furthermore, this model benefits from a broad frequency range and a moderately long receptive field size of 420~ms . Finally, the dilated convolutional model generalizes very well to unseen data, which is interesting for applicability in hearing assessment. 

\section*{Acknowledgments}
The authors thank Amelie Algoet, Jolien Smeulders, Lore Kerkhofs, Sara Peeters, Merel Dillen, Ilham Gamgami, Amber Verhoeven and Wendy Verheijen for their help with data collection.

\bibliography{refs, strings}

\begin{thebibliography}{10}

\bibitem{ding_emergence_2012}
N.~Ding and J.~Z. Simon, ``Emergence of neural encoding of auditory objects
  while listening to competing speakers,'' {\em Proceedings of the National
  Academy of Sciences}, vol.~109, pp.~11854--11859, July 2012.

\bibitem{vanthornhout_speech_2018}
J.~Vanthornhout, L.~Decruy, J.~Wouters, J.~Z. Simon, and T.~Francart, ``Speech
  {Intelligibility} {Predicted} from {Neural} {Entrainment} of the {Speech}
  {Envelope},'' {\em Journal of the Association for Research in
  Otolaryngology}, vol.~19, pp.~181--191, Apr. 2018.

\bibitem{lesenfants_data-driven_2019}
D.~Lesenfants, J.~Vanthornhout, E.~Verschueren, and T.~Francart, ``Data-driven
  spatial filtering for improved measurement of cortical tracking of multiple
  representations of speech,'' {\em Journal of Neural Engineering}, vol.~16,
  p.~066017, Oct. 2019.
\newblock Publisher: IOP Publishing.

\bibitem{crosse_multivariate_2016}
M.~J. Crosse, D.~Liberto, G.~M, A.~Bednar, and E.~C. Lalor, ``The
  {Multivariate} {Temporal} {Response} {Function} ({mTRF}) {Toolbox}: {A}
  {MATLAB} {Toolbox} for {Relating} {Neural} {Signals} to {Continuous}
  {Stimuli},'' {\em Frontiers in Human Neuroscience}, vol.~10, 2016.

\bibitem{Gillis2021.01.21.427550}
M.~Gillis, L.~Decruy, J.~Vanthornhout, and T.~Francart, ``Hearing loss is
  associated with delayed neural responses to continuous speech,'' {\em
  bioRxiv}, 2021.

\bibitem{de_cheveigne_decoding_2018}
A.~de~Cheveigné, D.~D.~E. Wong, G.~M. Di~Liberto, J.~Hjortkjær, M.~Slaney,
  and E.~Lalor, ``Decoding the auditory brain with canonical component
  analysis,'' {\em NeuroImage}, vol.~172, pp.~206--216, May 2018.

\bibitem{wong_accurate_2019}
D.~D.~E. Wong, G.~M.~D. Liberto, and A.~d. Cheveigné, ``Accurate {Modeling} of
  {Brain} {Responses} to {Speech},'' {\em bioRxiv}, p.~509307, July 2019.

\bibitem{akbari_towards_2019}
H.~Akbari, B.~Khalighinejad, J.~L. Herrero, A.~D. Mehta, and N.~Mesgarani,
  ``Towards reconstructing intelligible speech from the human auditory
  cortex,'' {\em Scientific Reports}, vol.~9, pp.~1--12, Jan. 2019.

\bibitem{iotzov_eeg_2019}
I.~Iotzov and L.~C. Parra, ``{EEG} can predict speech intelligibility,'' {\em
  Journal of Neural Engineering}, vol.~16, p.~036008, Mar. 2019.

\bibitem{lesenfants_predicting_2019}
D.~Lesenfants, J.~Vanthornhout, E.~Verschueren, L.~Decruy, and T.~Francart,
  ``Predicting individual speech intelligibility from the cortical tracking of
  acoustic- and phonetic-level speech representations,'' {\em Hearing
  Research}, vol.~380, pp.~1--9, Sept. 2019.
\newblock Publisher: Elsevier.

\bibitem{verschueren_neural_2019}
E.~Verschueren, B.~Somers, and T.~Francart, ``Neural envelope tracking as a
  measure of speech understanding in cochlear implant users,'' {\em Hearing
  Research}, vol.~373, pp.~23--31, Mar. 2019.

\bibitem{ciccarelli_comparison_2018}
G.~Ciccarelli, M.~Nolan, J.~Perricone, P.~Calamia, S.~Haro, J.~O'Sullivan,
  N.~Mesgarani, T.~Quatieri, and C.~Smalt, ``Comparison of {Two}-{Talker}
  {Attention} {Decoding} from {EEG} with {Nonlinear} {Neural} {Networks} and
  {Linear} {Methods},'' {\em bioRxiv}, p.~504522, Dec. 2018.

\bibitem{yang_speech_2015}
M.~Yang, S.~A. Sheth, C.~A. Schevon, G.~M.~M. II, and N.~Mesgarani, ``Speech
  {Reconstruction} from {Human} {Auditory} {Cortex} with {Deep} {Neural}
  {Networks},'' p.~5, 2015.

\bibitem{deckers_eeg-based_2018}
S.~Vandecappelle, L.~Deckers, N.~Das, A.~H. Ansari, A.~Bertrand, and
  T.~Francart, ``{EEG}-based detection of the locus of auditory attention with
  convolutional neural networks,'' {\em eLife}, vol.~10, p.~e56481, Apr. 2021.
\newblock Publisher: eLife Sciences Publications, Ltd.

\bibitem{geirnaert_neuro-steered_2020}
S.~Geirnaert, S.~Vandecappelle, E.~Alickovic, A.~de~Cheveigne, E.~Lalor, B.~T.
  Meyer, S.~Miran, T.~Francart, and A.~Bertrand, ``Electroencephalography-based
  auditory attention decoding: Toward neurosteered hearing devices,'' {\em IEEE
  Signal Processing Magazine}, vol.~38, no.~4, pp.~89--102, 2021.

\bibitem{taillez_machine_2017}
T.~de~Taillez, B.~Kollmeier, and B.~T. Meyer, ``Machine learning for decoding
  listeners’ attention from electroencephalography evoked by continuous
  speech,'' {\em European Journal of Neuroscience}, vol.~0, no.~0, 2017.

\bibitem{osullivan_attentional_2015}
J.~A. O'Sullivan, A.~J. Power, N.~Mesgarani, S.~Rajaram, J.~J. Foxe, B.~G.
  Shinn-Cunningham, M.~Slaney, S.~A. Shamma, and E.~C. Lalor, ``Attentional
  {Selection} in a {Cocktail} {Party} {Environment} {Can} {Be} {Decoded} from
  {Single}-{Trial} {EEG},'' {\em Cerebral Cortex}, vol.~25, pp.~1697--1706,
  July 2015.

\bibitem{10.1088/1741-2552/abf771}
A.~de~Cheveigné, M.~Slaney, S.~Fuglsang, and J.~Hjortkjaer, ``Auditory
  stimulus-response modeling with a match-mismatch task,'' {\em Journal of
  Neural Engineering}, 2021.

\bibitem{accou_modeling_2020}
B.~Accou, M.~J. Monesi, and J.~Montoya, ``Modeling the relationship between
  acoustic stimulus and {EEG} with a dilated convolutional neural network,'' in
  {\em 28th {European} {Signal} {Processing} {Conference} ({EUSIPCO} 2020)
  {EUSIPCO} 2020}, (Amsterdam), p.~1175, 2021.

\bibitem{oord_wavenet:_2016}
A.~v.~d. Oord, S.~Dieleman, H.~Zen, K.~Simonyan, O.~Vinyals, A.~Graves,
  N.~Kalchbrenner, A.~Senior, and K.~Kavukcuoglu, ``{WaveNet}: {A} {Generative}
  {Model} for {Raw} {Audio},'' {\em arXiv:1609.03499 [cs]}, Sept. 2016.
\newblock arXiv: 1609.03499.

\bibitem{luts_development_2014}
H.~Luts, S.~Jansen, W.~Dreschler, and J.~Wouters, ``Development and normative
  data for the {Flemish}/{Dutch} {Matrix} test,'' 2014.

\bibitem{decruy_self-assessed_2018}
L.~Decruy, N.~Das, E.~Verschueren, and T.~Francart, ``The {Self}-{Assessed}
  {Békesy} {Procedure}: {Validation} of a {Method} to {Measure}
  {Intelligibility} of {Connected} {Discourse}.,'' {\em Trends in Hearing},
  vol.~22, pp.~1--13, Oct. 2018.
\newblock Publisher: SAGE Publications UK and US.

\bibitem{francart_apex_2008}
T.~Francart, A.~van Wieringen, and J.~Wouters, ``{APEX} 3: a multi-purpose test
  platform for auditory psychophysical experiments,'' {\em Journal of
  Neuroscience Methods}, vol.~172, pp.~283--293, July 2008.

\bibitem{somers_generic_2018}
B.~Somers, T.~Francart, and A.~Bertrand, ``A generic {EEG} artifact removal
  algorithm based on the multi-channel {Wiener} filter,'' {\em Journal of
  Neural Engineering}, vol.~15, p.~036007, Feb. 2018.

\bibitem{sondergaard_linear_2012}
P.~L. Søndergaard, B.~Torrésani, and P.~Balazs, ``The linear time frequency
  analysis toolbox,'' {\em International Journal of Wavelets, Multiresolution
  and Information Processing}, vol.~10, p.~1250032, May 2012.
\newblock Publisher: World Scientific Publishing Co.

\bibitem{sondergaard_auditory_2013}
P.~L. Søndergaard and P.~Majdak, ``The {Auditory} {Modeling} {Toolbox},'' in
  {\em The {Technology} of {Binaural} {Listening}} (J.~Blauert, ed.), Modern
  {Acoustics} and {Signal} {Processing}, pp.~33--56, Berlin, Heidelberg:
  Springer, 2013.

\bibitem{biesmans_auditory-inspired_2017}
W.~Biesmans, N.~Das, T.~Francart, and A.~Bertrand, ``Auditory-{Inspired}
  {Speech} {Envelope} {Extraction} {Methods} for {Improved} {EEG}-{Based}
  {Auditory} {Attention} {Detection} in a {Cocktail} {Party} {Scenario},'' {\em
  IEEE Transactions on Neural Systems and Rehabilitation Engineering}, vol.~25,
  pp.~402--412, May 2017.

\bibitem{monesi_lstm_2020}
M.~J. Monesi, B.~Accou, J.~Montoya-Martinez, T.~Francart, and H.~V. Hamme, ``An
  {LSTM} {Based} {Architecture} to {Relate} {Speech} {Stimulus} to {Eeg},'' in
  {\em {ICASSP} 2020 - 2020 {IEEE} {International} {Conference} on {Acoustics},
  {Speech} and {Signal} {Processing} ({ICASSP})}, pp.~941--945, May 2020.
\newblock ISSN: 2379-190X.

\bibitem{martinabadi_and_tensorflow_2015}
{Mart\'in Abadi}, {Ashish Agarwal }, {Paul Barham }, {Eugene Brevdo }, {Zhifeng
  Chen }, {Craig Citro }, {Greg S. Corrado }, {Andy Davis }, {Jeffrey Dean },
  {Matthieu Devin }, {Sanjay Ghemawat }, {Ian Goodfellow }, {Andrew Harp },
  {Geoffrey Irving }, {Michael Isard }, {Yangqing Jia }, {Rafal Jozefowicz },
  {Lukasz Kaiser }, {Manjunath Kudlur }, {Josh Levenberg }, {Dan Man\'e },
  {Rajat Monga }, {Sherry Moore }, {Derek Murray }, {Chris Olah }, {Mike
  Schuster }, {Jonathon Shlens }, {Benoit Steiner }, {Ilya Sutskever }, {Kunal
  Talwar }, {Paul Tucker }, {Vincent Vanhoucke }, {Vijay Vasudevan }, {Fernanda
  Vi\'egas }, {Oriol Vinyals }, {Pete Warden }, {Martin Wattenberg }, {Martin
  Wicke }, {Yuan Yu }, and {Xiaoqiang Zheng}, ``Tensorflow: {Large}-{Scale}
  {Machine} {Learning} on {Heterogeneous} {Systems},'' 2015.

\bibitem{chollet_francois_keras_2015}
{Chollet, François}, ``Keras,'' 2015.

\bibitem{de_cheveigne_alain_noisetools_nodate}
{de Cheveigné, Alain}, ``{NoiseTools}.''

\bibitem{nair_rectified_2010}
V.~Nair and G.~E. Hinton, ``Rectified {Linear} {Units} {Improve} {Restricted}
  {Boltzmann} {Machines},'' p.~8, 2010.

\bibitem{bates_lme4_2021}
D.~Bates, M.~Maechler, B.~Bolker~[aut, cre, S.~Walker, R.~H.~B. Christensen,
  H.~Singmann, B.~Dai, F.~Scheipl, G.~Grothendieck, P.~Green, J.~Fox, A.~Bauer,
  and P.~N. K. s. c.~o. simulate.formula), ``lme4: {Linear} {Mixed}-{Effects}
  {Models} using '{Eigen}' and {S4},'' June 2021.

\bibitem{kuznetsova_lmertest_2020}
A.~Kuznetsova, P.~B. Brockhoff, R.~H.~B. Christensen, and S.~P. Jensen,
  ``{lmerTest}: {Tests} in {Linear} {Mixed} {Effects} {Models},'' Oct. 2020.

\bibitem{Team2014RAL}
R.~Team, ``R: A language and environment for statistical computing.,'' {\em
  MSOR connections}, vol.~1, 2014.

\bibitem{lenth_emmeans_2021}
R.~V. Lenth, P.~Buerkner, M.~Herve, J.~Love, H.~Riebl, and H.~Singmann,
  ``emmeans: {Estimated} {Marginal} {Means}, aka {Least}-{Squares} {Means},''
  Aug. 2021.

\bibitem{Crosse14195}
M.~J. Crosse, J.~S. Butler, and E.~C. Lalor, ``Congruent visual speech enhances
  cortical entrainment to continuous auditory speech in noise-free
  conditions,'' {\em Journal of Neuroscience}, vol.~35, no.~42,
  pp.~14195--14204, 2015.

\bibitem{ding_cortical_2014}
N.~Ding and J.~Z. Simon, ``Cortical entrainment to continuous speech:
  functional roles and interpretations,'' {\em Frontiers in Human
  Neuroscience}, vol.~8, 2014.
\newblock Publisher: Frontiers.

\bibitem{doelling_acoustic_2014}
K.~B. Doelling, L.~H. Arnal, O.~Ghitza, and D.~Poeppel, ``Acoustic landmarks
  drive delta–theta oscillations to enable speech comprehension by
  facilitating perceptual parsing,'' {\em NeuroImage}, vol.~85, pp.~761--768,
  Jan. 2014.

\bibitem{woodfield_role_2010}
A.~Woodfield and M.~A. Akeroyd, ``The role of segmentation difficulties in
  speech-in-speech understanding in older and hearing-impaired adults,'' {\em
  The Journal of the Acoustical Society of America}, vol.~128, pp.~EL26--EL31,
  June 2010.
\newblock Publisher: Acoustical Society of America.

\bibitem{diliberto_low-frequency_2015}
G.~M. Di Liberto, J.~A. O’Sullivan, and E.~C. Lalor, ``Low-{Frequency}
  {Cortical} {Entrainment} to {Speech} {Reflects} {Phoneme}-{Level}
  {Processing},'' {\em Current Biology}, vol.~25, pp.~2457--2465, Oct. 2015.
\newblock Publisher: Elsevier.

\bibitem{virtanen_scipy_2020}
P.~Virtanen, R.~Gommers, T.~E. Oliphant, M.~Haberland, T.~Reddy, D.~Cournapeau,
  E.~Burovski, P.~Peterson, W.~Weckesser, J.~Bright, S.~J. van~der Walt,
  M.~Brett, J.~Wilson, K.~J. Millman, N.~Mayorov, A.~R.~J. Nelson, E.~Jones,
  R.~Kern, E.~Larson, C.~J. Carey, I.~Polat, Y.~Feng, E.~W. Moore,
  J.~VanderPlas, D.~Laxalde, J.~Perktold, R.~Cimrman, I.~Henriksen, E.~A.
  Quintero, C.~R. Harris, A.~M. Archibald, A.~H. Ribeiro, F.~Pedregosa, P.~van
  Mulbregt, and {SciPy 1.0 Contributors}, ``{SciPy} 1.0: fundamental algorithms
  for scientific computing in {Python},'' {\em Nature Methods}, vol.~17, no.~3,
  pp.~261--272, 2020.

\bibitem{ding_neural_2011}
N.~Ding and J.~Z. Simon, ``Neural coding of continuous speech in auditory
  cortex during monaural and dichotic listening,'' {\em Journal of
  Neurophysiology}, vol.~107, pp.~78--89, Oct. 2011.
\newblock Publisher: American Physiological Society.

\bibitem{keshishian_estimating_2020}
M.~Keshishian, H.~Akbari, B.~Khalighinejad, J.~L. Herrero, A.~D. Mehta, and
  N.~Mesgarani, ``Estimating and interpreting nonlinear receptive field of
  sensory neural responses with deep neural network models,'' {\em eLife},
  vol.~9, p.~e53445, June 2020.
\newblock Publisher: eLife Sciences Publications, Ltd.

\bibitem{broderick_electrophysiological_2018}
M.~P. Broderick, A.~J. Anderson, G.~M. Di~Liberto, M.~J. Crosse, and E.~C.
  Lalor, ``Electrophysiological {Correlates} of {Semantic} {Dissimilarity}
  {Reflect} the {Comprehension} of {Natural}, {Narrative} {Speech},'' {\em
  Current Biology}, vol.~28, pp.~803--809.e3, Mar. 2018.

\end{thebibliography}

\clearpage

\vfill\null
\appendices%
\section{Sigmoid fit examples}

\begin{figure}[h]
    \centering
    \includegraphics[width=0.5\textwidth]{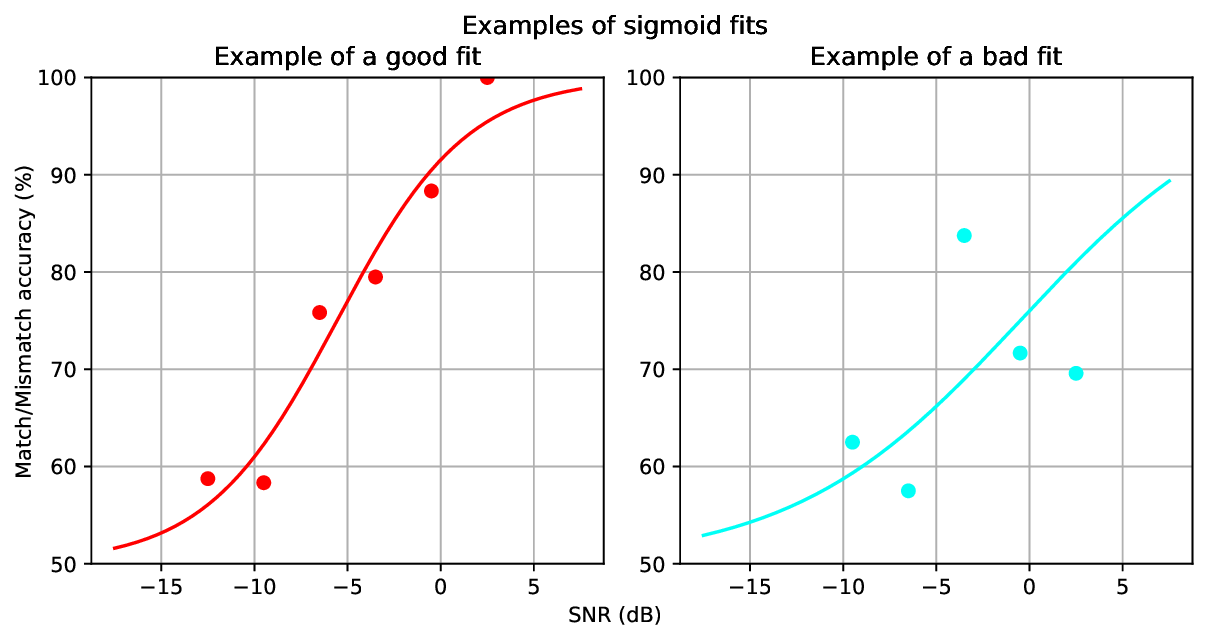}
    \caption{An example of a good sigmoid fit (left) and a discarded, bad sigmoid fit (right).}
    \label{fig:sigmoid_fits}
\end{figure}

\end{document}